\documentclass[12pt]{article}

\usepackage[T2A,T1]{fontenc}
\usepackage[utf8]{inputenc}
\usepackage[russian,english]{babel}

\pdfoutput=1

\usepackage{color}
\usepackage{graphicx}
\usepackage{amsmath}
\usepackage{amssymb}
\usepackage{xspace}
\usepackage[small]{subfigure}
\usepackage[hyperfootnotes=false,hidelinks]{hyperref}

\newlength{\fighskip} \fighskip=2pt
\newlength{\figvskip} \figvskip=3pt

\usepackage{mciteplus} 
\usepackage{dcolumn}
\usepackage{bm}
\usepackage{verbatim}
\usepackage{amscd}
\usepackage{amsfonts}
\usepackage{setspace}
\usepackage{amsthm}
\usepackage{enumerate}
\usepackage{mathtools}

\newcommand{\blangle}{\bigl\langle}
\newcommand{\brangle}{\bigr\rangle}
\newcommand{\dlangle}{\langle\kern-1.5pt\langle}
\newcommand{\drangle}{\rangle\kern-1.5pt\rangle}
\newcommand{\bdlangle}{\blangle\kern-3pt\blangle}
\newcommand{\bdrangle}{\brangle\kern-3pt\brangle}

\newcommand{\bra}[1]{\langle {#1} \vert}
\newcommand{\ket}[1]{\vert {#1} \rangle}

\title{Not even 6 dB: Gaussian quantum illumination in thermal background}
\author{T.J.\,Volkoff\thanks{\texttt{volkoff@lanl.gov}}\\ {\small \textit{Theoretical Division, Los Alamos National Laboratory, Los Alamos, NM, USA.}}}
\date{}

\begin{document}

\maketitle

\begin{abstract}
In analyses of target detection with Gaussian state transmitters in a thermal background, the thermal occupation is taken to depend on the target reflectivity in a way which simplifies the analysis of the symmetric quantum hypothesis testing problem. However, this assumption precludes comparison of target detection performance between an arbitrary transmitter and a vacuum state transmitter, i.e., ``detection without illumination'', which is relevant in a bright thermal background because a target can be detected by its optical shadow or some other perturbation of the background. Using a target-agnostic thermal environment leads to the result that the oft-claimed 6 dB possible reduction in the quantum Chernoff exponent for a two-mode squeezed vacuum transmitter over a coherent state transmitter in high-occupation thermal background is an unachievable limiting value, only occurring in a limit in which the target detection problem is ill-posed. Further analyzing quantum illumination in a target-agnostic thermal environment shows that a weak single-mode squeezed transmitter performs worse than ``no illumination'', which is explained by the noise-increasing property of reflected low-intensity squeezed light.
\end{abstract}

%\end{titlepage}

%\tableofcontents
%\baselineskip=17.63pt

\section{Introduction}

Quantum illumination (QI) is a symmetric hypothesis testing problem in which a decision is made between ``target absent'' and ``target present'' after transmitting many copies of a quantum subsystem to the target \cite{lloyd,shapiro}. The full quantum system may have quantum memory registers which are not transmitted to the target. The model of the target includes a small reflectivity $\kappa \ll 1$ beamsplitter uniformly coupling the transmitted registers and corresponding thermal modes of energy $N_{B}$, and, in analyses of the purely information-theoretic QI problem, has no other interesting property like spatial or temporal dynamics, thermal or active emission, absorption, etc. Such properties are important for realistic analyses of QI problem involving target properties or specific receivers \cite{PhysRevA.103.062413,Jonsson2021QuantumR,PhysRevA.105.042412, lee2022,PhysRevApplied.20.014030,PhysRevA.80.052310}, but they are not considered in the present work. In Gaussian QI, the full state of the transmitted register $T$ and the quantum memory register $Q$ is a continuous-variable (CV) Gaussian state $\ket{\psi}_{TQ}$. One can, of course, consider alternative information-theoretic settings such as asymmetric hypothesis testing setting \cite{PhysRevLett.119.120501}, or first-photon radar \cite{PhysRevA.106.032415}, but the present work is concerned with the traditional setting.

For QI in an athermal environment (number of thermal environment photons $N_{B}=0$) and a target modeled by a quantum-limited attenuator channel, Nair's ``no-go'' theorem establishes that the optimal error probability for the optimal transmitter is 1/2 of the optimal error probability for a coherent state transmitter in the limit of small reflectivity, $\kappa \ll 1$ \cite{PhysRevA.84.032312}. The factor of 1/2 is merely an additive increase of the quantum Chernoff exponent. In contrast, in a bright thermal background ($N_{B}\gg 1$ per mode), and assuming a certain model of the target, it has been shown that a two-mode squeezed state (TMSS) transmitter with transmitted intensity $N_{S}$
attains about a 6 dB (specifically, a factor of exactly 4) smaller quantum Chernoff exponent than a coherent state transmitter $\ket{\alpha}_{T}$ with $\alpha = \sqrt{N_{S}}$. The supporting calculation is based on the comparison of the Bhattacharyya bounds on the optimal error probability, which is justified in the parameter domains considered. However, and most importantly, the thermal background mode is assumed to depend on the target's parameter $\kappa$ in such a way that, regardless of the value of $\kappa$, the thermal environment increases the noise in each quadrature of the reflected state by an additive amount $N_{B}+{1\over 2}$. The assumption is implemented by taking $N_{B}\mapsto {N_{B}\over 1-\kappa}$ after the beamsplitter transformation is made. Such a description of the target is still a bosonic Gaussian channel, but it is not a unitary beamsplitter. Beyond the mathematical convenience of this assumption, it is sometimes justified by the restriction that the target must be \textit{illuminated} for target detection to occur (a vacuum transmitter does not give a well-defined QI problem under this model of the target-- the states $\rho_{\kappa}$ and $\rho_{0}$ corresponding to the ``target present'' and ``target absent'' hypotheses, respectively, would be exactly the same.). The present work dispenses with this assumption (called the ``no passive signatures'' assumption in \cite{covdet}) and its ensuing mathematical convenience, instead taking the target to be defined in a simple way by the thermal attenuator channel involving a beamsplitter of reflectivity $\sqrt{\kappa}$ and an environment consisting of a bare thermal background $\rho_{\beta}:= \sum_{n=0}^{\infty}\left( {N_{B}\over N_{B}+1}\right)^{n}\ket{n}\bra{n}_{E}$, where $\beta:=\ln {1+N_{B}\over N_{B}}$ is the reciprocal of the effective temperature of the environment mode. Such a definition is a straightforward generalization of the quantum-limited attenuator channel used in the noiseless target detection problem-- just put a thermal state on the beamsplitter instead of vacuum. The present model of the target has a major consequence: the QI problem is well-posed even for vacuum transmitters because the quadrature noises are different between the ``target present'' and ``target absent'' hypotheses. As shown in Section \ref{sec:ooo}, the possibility of vacuum detection also leads to corrections to the coherent state transmitter QI error probability. The main results of the present work are in Sections \ref{sec:uuu}, \ref{sec:sss} which show, respectively, that nonclassical Gaussian transmitters can perform worse than coherent state transmitters, and that the oft-claimed 6 dB advantage of two-mode squeezed state transmitters over classical transmitters for Gaussian QI in a bright thermal background is strictly unachievable.

It has been noted that if the rescaling $N_{B}\mapsto {N_{B}\over 1-\kappa}$ is not made, a receiver can be designed based on double homodyne detection \cite{PhysRevResearch.3.013006} that, for $N_{B}\gg 1$, outperforms previously proposed non-passive receivers requiring amplification or phase-conjugation \cite{PhysRevA.80.052310}. Therefore, assuming this rescaling can lead to  changes in optimal detection strategy. In the related task of \textit{estimation} of the reflectivity parameter $\kappa$ with quantum state probes, the quantum Fisher information (QFI) is the relevant quantity determining an achievable lower bound for the error of an estimate of $\kappa$ (quantum Cram\'{e}r-Rao theorem \cite{PhysRevLett.72.3439}). Due to the fact that the quantum Fisher information can be interpreted as the metric tensor associated with the Bures distance \cite{PhysRevLett.72.3439}, Fuchs-van de Graaf inequalities allow to derive QFI-based bounds on the quantum Chernoff exponent for the $\rho_{0}$ versus $\rho_{\kappa}$ symmetric hypothesis testing problem, which are valid for $\kappa \ll 1$ \cite{PhysRevLett.118.100502,PhysRevLett.118.070803,Jonsson_2022,leeng}.  In this context, it has been shown that the advantage of a two-mode squeezed vacuum probe over a coherent state probe in the many-copy limit and in the presence of a thermal environment occurs in a restricted parameter domain when the assumption $N_{B}\mapsto {N_{B}\over 1-\kappa}$ is made, compared to when it is not made \cite{Jonsson_2022}.  

\section{Background and methods}
The physical motivations for QI and the relation of the QI problem to classical radar detection can be found in the seminal paper Ref.\cite{lloyd}. In Section \ref{sec:ooo}, we show how the time-bandwidth pulse structure of transmitter states assumed in \cite{lloyd} is related to the two-mode structure of the Gaussian states considered in the present work.
 
Our convention for the formalism for CV Gaussian states follows Ref.\cite{holevo}. Specifically, $R=(q_{1},p_{1},\ldots,q_{M},p_{M})$ is the row vector of canonical operators ($[q_{j},p_{j'}]=i\delta_{j,j'}$)  of an $M$-mode CV system, $a_{j}={q_{j}+ip_{j}\over\sqrt{2}}$ is an annihilation operator, and we use the symplectic form $\Delta$ on the phase space $\mathbb{R}^{2M}$  associated with matrix $\begin{pmatrix}
0&1\\-1&0
\end{pmatrix}^{\oplus M}$ which allows to define the symplectic group $Sp(2M,\mathbb{R})$ as those $2M\times 2M$ matrices satisfying $S^{T}\Delta S=\Delta$. The covariance matrix of a Gaussian state $\rho$ is defined by $\Sigma_{\rho}:= \langle (R - m_{\rho})^{T}\circ (R - m_{\rho})\rangle_{\rho}$, where $\circ$ is the Jordan product and $m_{\rho}:=\langle R \rangle_{\rho}$ is the mean vector (of phase space displacement). The structure of the covariance matrix is determined by its symplectic diagonalization, which identifies $M$ normal modes of the system. This and further properties of CV Gaussian states are discussed in Refs.\cite{holevo,serafini}.

In QI, the task is to distinguish optimally between $\rho_{0}^{\otimes N}$ and $\rho_{\kappa}^{\otimes N}$, $\kappa\neq 0$ as $N\rightarrow \infty$ with these hypotheses having equal prior probability (i.e., symmetric hypothesis testing). More details about symmetric hypothesis testing and the quantum Chernoff bound appear in \cite{PhysRevLett.98.160501}. For the purpose of the present work, which is focused on evaluation of the asymptotic error probability for this task, one needs to compute the quantum Chernoff exponent $\xi$ on the set Gaussian states \cite{PhysRevA.78.012331}, where
\begin{align}
\xi(\rho_{0},\rho_{\kappa})&:= -\log\left(\inf_{0\le s\le 1}Q_{s}(\rho_{0},\rho_{\kappa})\right) \nonumber \\
Q_{s}(\rho_{0},\rho_{\kappa})&:= \text{tr}\rho_{0}^{s}\rho_{\kappa}^{1-s},
\label{eqn:qce}
\end{align}
and use the fact that $p_{\text{err}} \sim {1\over 2}e^{-N\xi}$ as $N\rightarrow \infty$ \cite{PhysRevLett.98.160501}. The quantity $2(1-Q_{s})$ is a special case of a relative $g$-entropy, or Petz-R\'{e}nyi relative quasi-entropy, or a quantum $g$-divergence corresponding to the operator convex function $g(t)=2(1-t^{s})$, $0\le s\le 1$ \cite{ohyapetz,petzvonneumann,bhatia,Wilde_2018, 10.1063/1.5039973}. The specific quantity $2(1-Q_{1/2})$ is the quantum analogue of the Hellinger distance, with $Q_{1/2}$ (referred to as the quantum affinity \cite{holevoquasiequiv,PhysRevA.69.032106}) appearing in the Bhattacharyya upper bound to the optimal discrimination error, $p_{\text{err}} \le {1\over 2}Q_{1/2}^{N}$ (in classical statistics, \textit{affinity} is another name for the Bhattacharyya coefficient). Note that by invoking the one-to-one correspondence \cite{ruskai} between relative $g$-entropies (specifically, $g(t)=2(1-\sqrt{t})$) and monotone Riemannian metrics on the quantum state space  (specifically, the one defined by the Chentsov-Morozova function $c(x,y)={2\over (\sqrt{x}+\sqrt{y})^{2}}$, sometimes called the Wigner-Yanase monotone metric \cite{PhysRevA.91.042330}), one can obtain the $O(\kappa^{2})$ approximation to $\inf_{0\le s\le 1}Q_{s}(\rho_{0},\rho_{\kappa})$  \cite{10.1063/1.1598279,lee2023}. Note that, using properties of the quantum fidelity and number-diagonal transmitter modes \cite{PhysRevA.84.032312}, an upper bound on $\xi(\rho_{0},\rho_{\kappa})$ has been derived which is valid for all transmitter states and for all scalings of the thermal environment noise parameter, including the ``passive signature'' and ``no passive signature'' settings \cite{covdet}.

Like Ref.\cite{PhysRevLett.101.253601}, and most other analyses of QI, we will make the small reflectivity assumption $0< \kappa \ll 1$ for the ``target present'' hypothesis throughout this work.
%\begin{figure}[h]
%    \centering
%    \includegraphics[scale=0.5]{fd}
%    \caption{...}
%    \label{fig:fd}
%\end{figure}
%
%
%
%
%\begin{figure*}
%    \centering
%    \includegraphics[scale=0.5]{vis}
%    \caption{...}
%    \label{fig:f1}
%\end{figure*}

\section{Gaussian quantum illumination with coherent states\label{sec:ooo}}

In Ref.\cite{lloyd}, the time-bandwidth product is used to define the $M$ modes of the transmitter register $T$ (or $2M$ modes of the transmitter-plus-quantum-memory register $TQ$ in the case of an entangled transmitter). Each of the $M$ modes, which we associate with annihilation operator $a_{j}$, $j\in [M]$, is assumed to reflect from the target with probability $\kappa$. Therefore, the target is modeled as a beamsplitter ($\theta:= \cos^{-1}\sqrt{\kappa}$, $\theta \in [0,{\pi\over 2}]$)
\begin{equation}
U_{\kappa}:=e^{-\theta\sum_{j=0}^{M-1}\left(a_{j}b_{j}^{\dagger}-h.c.\right)}
\end{equation}
uniformly interfering each $a_{j}$ with its corresponding mode $b_{j}$ of the environment register $E$. Each environment mode is assumed thermal with intensity $N_{B}$. For a coherent state transmitter, we will consider $N_{S}/M$ photons in expectation in each temporal mode $j$ and take zero phase of each coherent state without loss of generality, leading to the parametrized model
\begin{equation}
    \rho_{\kappa}=\text{tr}_{E} \left[ U_{\kappa}(D_{\tilde{0}}(\sqrt{N_{S}})\otimes \mathbb{I}_{E})\ket{\text{VAC}}\bra{\text{VAC}}_{T}\otimes \rho_{\beta}^{\otimes M}(D_{\tilde{0}}(-\sqrt{N_{S}})\otimes \mathbb{I}_{E}) U_{\kappa}^{\dagger}\right]
    \label{eqn:sss}
\end{equation}
where the Fourier mode $\tilde{j}$ of register $T$ is associated with annihilation operator
$\tilde{a}_{j}:={1\over \sqrt{M}}\sum_{k=0}^{M-1}e^{2\pi i jk\over M}a_{k}$, and a CV displacement $D_{\tilde{j}}(z):=e^{z\tilde{a}_{j}^{\dagger}-\overline{z}\tilde{a}_{j}}$, $z \in \mathbb{C}$. Written in terms of Fourier modes, both $U_{\kappa}$ and $\rho_{\beta}^{\otimes M}$ are unchanged in form, so $U_{\kappa}$ couples the $\tilde{a}_{0}$ mode to (and only to) the $\tilde{b}_{0}^{\dagger}$ mode. Taking the partial trace over all but the $\tilde{b}_{0}$ mode of $E$ leaves a single remaining partial trace to be carried out over the $\tilde{b}_{0}$ mode:
\begin{equation}
\rho_{\kappa}= \text{tr}_{E_{\tilde{0}}}\left[ e^{-\theta\left(\tilde{a}_{0}\tilde{b}_{0}^{\dagger} - h.c.\right)}D_{\tilde{0}}(\sqrt{N_{S}})\ket{\text{VAC}}\bra{\text{VAC}}_{T}D_{\tilde{0}}(\sqrt{N_{S}})^{\dagger}\otimes \rho_{\beta} e^{\theta\left(\tilde{a}_{0}\tilde{b}_{0}^{\dagger} - h.c.\right)} \right]
\label{eqn:csbs}
\end{equation}
which shows that the non-vacuum support of $\rho_{\kappa}$ is described by a single-mode bosonic Gaussian channel (specifically, the thermal attenuator channel \cite{holevo}) acting on the single-mode coherent state $D_{\tilde{0}}(\sqrt{N_{S}})\ket{0}_{T_{\tilde{0}}}$ of expected energy $N_{S}$. The quantum channel that describes the reflection process in Ref.\cite{PhysRevLett.101.253601} with the rescaling $N_{B}\mapsto {N_{B}\over 1-\kappa}$ can also be written as a bosonic Gaussian channel, but actually involves amplification, which we view as an exotic type of target, not the passive reflective target of the originally envisioned QI task. When analyzing Gaussian quantum illumination in a thermal background, we thus restrict to a single-mode transmitter mode $T$ (potentially entangled with a single-mode memory $Q$), and a single-mode thermal environment $E$, no longer explicitly identifying Fourier modes of a register defined by the time-bandwidth product. This simplification is also made in Ref.\cite{PhysRevLett.101.253601}.

The covariance matrix of $\rho_{\kappa}$ in (\ref{eqn:csbs}) (i.e., the state under hypothesis ``target present'') is given by
\begin{equation}
\Sigma_{\rho_{\kappa}}={1\over 2}\begin{pmatrix}
1+2N_{B}(1-\kappa)&0\\0&1+2N_{B}(1-\kappa)
\end{pmatrix},
\end{equation}
its mean vector is given by $m_{\rho_{\kappa}}=(\sqrt{2\kappa N_{S}},0)$. The covariance matrix of $\rho_{0}$ (the state under hypothesis ``target absent'') is $\Sigma_{\rho_{0}}$, and the mean vector is $m_{\rho_{0}}$. It is important to notice that even if the transmitter register $T$ were vacuum, i.e., $N_{S}=0$, so that $m_{\rho_{0}}=m_{\rho_{\kappa}}=0$, the hypotheses $\rho_{0}$ and $\rho_{\kappa}$ could still be distinguished due to the different power in the quadrature noise. This comparison is not possible using the method of Ref.\cite{PhysRevLett.101.253601}. Physically, one can consider the difference of the present approach from Ref.\cite{PhysRevLett.101.253601} as fully taking into account the optical shadow of the target cast by the background thermal light. More generally, the background thermal light can be considered as a non-cooperative transmitter (i.e., ``illuminator of opportunity'' \cite{LIU201532}) for target detection in a quantum passive radar setup. Classical passive radar  is an active area of radar research \cite{pr}, and several classical hypothesis testing settings have been considered, including passive bistatic radar  \cite{pr} and multistatic radar \cite{9316305}, multitarget passive radar \cite{7849236}, and passive radar with direct path interference \cite{LIU201532,electronics12020433}.

 We also emphasize that quantum illumination with a vacuum transmitter differs from general ghost imaging, in which part of a correlated source interacts with a target and the target is imaged by measuring intensity correlations of the full output \cite{RevModPhys.94.025007,Ragy2012}. Although reflection from the target does cause the reflected signal mode to become correlated with background thermal light (even if the signal mode is vacuum, as in the beamsplitter appearing in thermal ghost imaging \cite{PhysRevLett.93.093602}), the detection protocols considered in standard QI settings are constrained to occur on the reflected signal modes. If a global measurement on the correlated reflected signal mode and thermal environment mode were allowed, then the principal difference would be that QI is usually concerned with optimal measurements, whereas thermal ghost imaging usually measures spatial correlations of the intensity.

Because $\rho_{0}$ and $\rho_{\kappa}$ are Gaussian states, Theorem 2 of Ref.\cite{PhysRevA.78.012331} or Theorem 18 of Ref.\cite{slw} can be used according to convenience to compute $Q_{s}$ in (\ref{eqn:qce}). In the latter reference, $Q_{s}$ depends in a more or less simple way on the mean vectors and covariance matrices of $\rho_{0}$ and $\rho_{\kappa}$, whereas the former reference computes $Q_{s}$ from data of the symplectic diagonalizations (Williamson theorem) applied to $\Sigma_{\rho_{0}}$ and $\Sigma_{\rho_{\kappa}}$.  The vector of canonical operators in Ref.\cite{PhysRevA.78.012331} is given by $\sqrt{2}R$, so their symplectic eigenvalues are multiplied by 2 compared to the ones in the present work, and their mean vectors are multiplied by $\sqrt{2}$ compared to the ones in the present work. Using the definitions from Ref.\cite{PhysRevA.78.012331} for the real-valued functions $G_{p}(x)$ and $\Lambda_{p}(x)$, where $x\ge 1$, $p\ge 0$, and noting that $\Sigma_{\rho_{0}}$ and $\Sigma_{\rho_{\kappa}}$ are symplectically diagonalized by the $2\times 2$ identity matrix $\mathbb{I}_{2}$,  one finds that
\begin{align}
Q_{s} &= {2G_{s}(2N_{B}+1)G_{1-s}(2(1-\kappa)N_{B}+1)e^{-{2\kappa N_{S}\over\Lambda_{s}(2N_{B}+1)+\Lambda_{1-s}(2(1-\kappa)N_{B}+1)} }\over \sqrt{ \text{det} \begin{pmatrix}
\Lambda_{s}(2N_{B}+1)+\Lambda_{1-s}(2(1-\kappa)N_{B}+1) & 0 \\ 0 & \Lambda_{s}(2N_{B}+1)+\Lambda_{1-s}(2(1-\kappa)N_{B}+1)
\end{pmatrix} }} \nonumber \\
&= {\exp\left[ -4\kappa N_{S}\left( {(N_{B}+1)^{s}((1-\kappa)N_{B}+1)^{1-s} - N_{B}(1-\kappa)^{1-s} \over ((N_{B}+1)^{s}-N_{B}^{s})(((1-\kappa)N_{B}+1)^{1-s}-(1-\kappa)^{s}N_{B}^{1-s})} \right)\right]\over (1+N_{B})\left( 1-{\kappa N_{B}\over 1+N_{B}} \right)^{1-s} - N_{B}(1-\kappa)^{1-s}}.
\label{eqn:uuu}
\end{align}
If the minimum over $s$ that defines the quantum Chernoff exponent $\xi$ in (\ref{eqn:qce}) were achieved at $s=1/2$ in some limit of the parameters $\kappa$ and $N_{B}$, it would imply that the Bhattacharyya upper bound ${1\over 2}Q_{1/2}^{N}$ to the optimal error probability is tight. For $N_{S}\gg 0$, the minimal value of $Q_{s}$ is approximately determined by the minimal value of  $\Lambda_{s}(2N_{B}+1)+\Lambda_{1-s}(2(1-\kappa)N_{B}+1)$ due to the exponentially vanishing factor. With $N_{B}\gg 0$, this has the following expansion
\begin{equation}
\Lambda_{s}(2N_{B}+1)+\Lambda_{1-s}(2(1-\kappa)N_{B}+1) = {2N_{B}(1-\kappa s) +1\over s(1-s)} + O(1/N_{B})
\label{eqn:sumsum}
\end{equation}
and the leading order term is minimized at $s={1\over 2}+{\kappa N_{B}\over 4(2N_{B}+1)} + o(\kappa)$ for $\kappa \ll 1$. Therefore, for $\kappa \ll 1$ and $N_{B}\gg 0$, the Bhattacharyya bound describes the optimal error probability well. By contrast, for $N_{B}\ll 1$ the exponential factor has no dependence on $s$, as can be seen from the asymptotic
\begin{equation}
\Lambda_{s}(2N_{B}+1)+\Lambda_{1-s}(2(1-\kappa)N_{B}+1)= 2+2(N_{B}^{s}+N_{B}^{1-s}) + O(N_{B})+O(N_{B}^{1-s}\kappa).
\end{equation} One then looks to maximize the denominator of (\ref{eqn:uuu}) to minimize $Q_{s}$. Expanding the derivative of this denominator to first order in $N_{B}$ and third order in $\kappa$, one obtains a quadratic equation for the critical point with solution ${1\over 2}+{\kappa\over 24}+o(\kappa^{2})$.  Therefore, for $N_{B}\ll 1$ and $\kappa \ll 1$, the Bhattacharyya bound again describes the optimal error probabilfity well regardless of $N_{S}$.

We now examine the Bhattacharyya bounds in the parameter domains described above. For $N_{B}\gg 0$, the Bhattacharyya bound
\begin{align}
{1\over 2}Q_{1/2}^{N}&={1\over 2}\left( 1-{(N_{B}-1)\kappa^{2}\over 8N_{B}} +O({\kappa^{2}\over N_{B}^{2}}) \right)^{N}e^{-{N\kappa N_{S}\over 2((2-\kappa)N_{B} +1) + O(N_{B}^{-1})}} \nonumber \\
&\le {1\over 2}e^{-N \left( {\kappa^{2}(N_{B}-1)\over 8N_{B}} + {\kappa N_{S}\over 2((2-\kappa)N_{B} +1)} \right)}
\label{eqn:b1}
\end{align}
holds for sufficiently large $N_{B}$, where the $N_{S}$ dependent exponential factor is found by taking $s=1/2$ in (\ref{eqn:sumsum}); this factor agrees with the result of Ref.\cite{PhysRevLett.101.253601} in the limit $N_{B}\rightarrow \infty$, $\kappa\rightarrow 0$. The first term in the exponent can be considered as a reduction of optimal error probability due to the vacuum contribution, and is absent from Ref.\cite{PhysRevLett.101.253601}. It is negligible compared to the second term in the limit $\kappa\rightarrow 0$. In the parameter domain $N_{B}\ll 1$, the Bhattacharyya bound is
\begin{align}
{1\over 2}Q_{1/2}^{N}&={1\over 2}\left( 1-{N_{B}\kappa^{2}\over 8} +O(N_{B}^{2}\kappa^{2}) \right)^{N}e^{-{N\kappa N_{S}\over 1+2\sqrt{N_{B}}+O(N_{B})+O(\kappa\sqrt{N_{B}})}} \nonumber \\
&\le {1\over 2}e^{-N \left( {N_{B}\kappa^{2}\over 8} + {\kappa N_{S}\over 1+2\sqrt{N_{B}}} \right)}
\label{eqn:b2}
\end{align}
where the inequality is valid for sufficiently small $N_{B}$. The vacuum contribution to (\ref{eqn:b2}) is shown only to indicate its existence. It is not only negligible as $N_{B}\rightarrow 0$ or $\kappa\rightarrow 0$, but for large transmitter intensity, for $N_{S}>1$ it is not even the dominant term of order 2 in $\kappa$ because $O(\kappa\sqrt{N_{B}})$ contributes in the denominator of the non-vacuum contribution. In the QI setting of the present work, the Bhattacharyya bounds (\ref{eqn:b1}), (\ref{eqn:b2})  are the fiducial values to which optical transmitters of intensity $N_{S}$ will be compared to in the respective large $N_{B}$ and small $N_{B}$ regimes.

Before moving on to consider nonclassical states as transmitters, we note that for $N_{B}=0$, coherent states minimize $p_{\text{err}}$ over all single-mode states of a fixed energy in the limit $\kappa \rightarrow 0$ \cite{PhysRevA.103.062413,lee2023}. This indicates that a nonzero thermal background $N_{B}>0$ is necessary for identifying an advantage of nonclassical transmitter states in the QI task.

\section{Gaussian quantum illumination with squeezed vacuum\label{sec:uuu}}

\begin{figure}[h]
    \centering
    \includegraphics[scale=0.9]{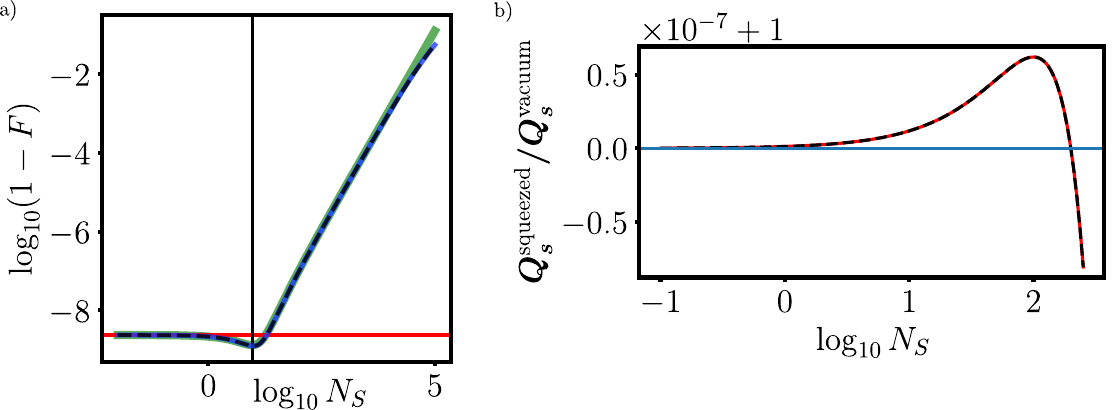}
    \caption{a) (Red) Fidelity $F\left(\text{tr}_{E}\left[ U_{\kappa}\ket{0}\bra{0}_{T}\otimes \rho_{\beta} U_{\kappa}^{\dagger}\right],\rho_{\beta} \right)$ with $\kappa = 10^{-4}$ and $N_{B}=20$. (Dashed black and underlying blue curve) Coinciding analytical and numerical fidelity $F\left(\text{tr}_{E}\left[ U_{\kappa}S(r)\ket{0}\bra{0}_{T}S(-r)\otimes \rho_{\beta} U_{\kappa}^{\dagger}\right],\rho_{\beta}\right)$ with $N_{S}=\sinh^{2}r$ for the same $\kappa = 10^{-4}$ and $N_{B}=20$. (Green)  Approximate fidelity from small $\kappa$ and large $N_{B}$ expansion of the analytical formula \cite{PhysRevLett.115.260501}. The vertical black line is the maximal value of the approximate analytical fidelity, $N_{S}={2N_{B}-3\over 4}$. b) (Red) Ratio of $Q_{s_{*}}:=\inf_{0\le s\le 1}Q_{s}$ of single-mode squeezed vacuum transmitter to $Q_{s_{*}}$ for vacuum transmitter, with varying $N_{S}$ for $N_{B}=200$, $\kappa=10^{-3}$; $s_{*}$ is the minimum identified by numerical minimization of $Q_{s}$ for each transmitter. (Black dashed) $Q_{1/2}$ for single-mode squeezed vacuum transmitter with same parameter values. Blue line is the value 1. A value of this ratio being greater than 1 indicates that the single-mode squeezed transmitter has a larger error probability in target detection than vacuum.}
    \label{fig:ttt}
\end{figure}

Gaussian QI with the $N_{B}$ rescaling of Ref.\cite{PhysRevLett.101.253601} was considered for a general single mode Gaussian state in Ref.\cite{lee2022}, with only a negligible $O(\kappa^{2})$ benefit of squeezing observed for the optimal error probability exponent quantified by signal-to-noise ratio of an on-off receiver or photon number-resolving receiver in the domain $\kappa \ll 1$, $N_{B}\gg 1$. The vacuum contributions to the optimal error probability derived in Section \ref{sec:ooo} imply that replacing $N_{B}\mapsto N_{B}/(1-\kappa)$ in the covariance matrix of the reflected Gaussian state leads to an underestimate of the success probability of QI. One consequence of our not carrying out this replacement is that it becomes meaningful to compare the QI performance of a vacuum transmitter to a squeezed vacuum transmitter. Although a squeezed vacuum transmitter indeed illuminates the target and, for sufficiently small $\kappa$, squeezed photons are reflected from the target, it is not obvious that a squeezed vacuum transmitter should result in a lower optimal error probability compared to bare vacuum (zero intensity transmitter) because the relation between quadrature noise anisotropy and detection is not clear. Fig. \ref{fig:ttt} compares the quantum fidelity (recall the quantum fidelity is defined by $F(\rho_{1},\rho_{2}):=\left( \text{tr}\left[ \sqrt{\sqrt{\rho_{1}}\rho_{2}\sqrt{\rho_{1}}} \right]\right)^{2}$) between reflected squeezed vacuum (``target present'') and a thermal state (over a range of $N_{S}$) to the fidelity between transmitted thermal state and a thermal state (no $N_{S}$ dependence because no photons are transmitted).
There is a domain of squeezing strengths for which the reflected squeezed state is more similar to the thermal environment than is the state resulting from sending no transmitter.  Quadrature noise provides some insight into this fact. One can note that taking a convex combination of the covariance matrices corresponding to slightly squeezed vacuum and thermal noise results in a quadrature ellipse that is, on angular average, less distinguishable from thermal noise than an isotropic contraction of the thermal noise covariance by a factor $1-\kappa$.  This fact motivates a counterintuitive hypothesis: there may be a domain of squeezing strengths for which a squeezed transmitter state is detrimental for QI compared to sending no transmitter. We prove this hypothesis by showing the validity of the Bhattacharyya bound in this case and comparing the bounds.

The covariance matrix of reflected squeezed vacuum transmitter $\rho_{\kappa}$ is given by
\begin{equation}
\Sigma_{\rho_{\kappa}}= {1\over 2}\begin{pmatrix}
\kappa e^{-2r}+(1-\kappa)(2N_{B}+1) & 0 \\ 0 & \kappa e^{2r} + (1-\kappa)(2N_{B}+1)
\end{pmatrix} = \nu_{1}SS^{T}
\label{eqn:sqcov}
\end{equation}
where $S$ is a $2\times 2$ symplectic matrix (which turns out to simply describe squeezing of a certain strength) and $\nu_{1}:= {1\over 2}\sqrt{(\kappa e^{2r}+(1-\kappa)(2N_{B}+1))(\kappa e^{-2r}+(1-\kappa)(2N_{B}+1))}$ is the symplectic eigenvalue. The mean vector of $\rho_{\kappa}$ is $(0,0)$. With (\ref{eqn:sqcov}) in hand, we aim to justify the statement that $s=1/2$ is a good approximation of the minimum of $Q_{s}$ when $\kappa \ll 1$. Given zero-mean, one-mode Gaussian states $\rho_{0}$, $\rho_{1}$ with covariance matrices having symplectic eigenvalues $\nu_{0}$, $\nu_{1}$ and symplectic diagonalizations $S_{0}$, $S_{1}$, respectively, introduce the functions
\begin{align}
F_{0}(s,\nu_{0},\nu_{1})&:= {\left( (2\nu_{0}+1)^{s}+(2\nu_{0}-1)^{s} \right) \left( (2\nu_{1}+1)^{1-s}-(2\nu_{1}-1)^{1-s} \right) \over 4} \nonumber \\
F_{1}(s,\nu_{0},\nu_{1})&:= F_{0}(1-s,\nu_{1},\nu_{0})
\end{align}
so that
\begin{equation}
Q_{s}= 2e^{-{1\over 2}\text{tr}\log \left[ F_{0}S_{0}S_{0}^{T}+ F_{1}S_{1}S_{1}^{T} \right]}.
\label{eqn:altq}
\end{equation}
For $\nu_{0},\nu_{1}\gg 1/2$, $F_{0}(s,\nu_{0},\nu_{1}) \sim (1-s)\nu_{0}^{s}\nu_{1}^{-s}$ and $F_{1}(s,\nu_{0},\nu_{1})\sim F_{0}(1-s,\nu_{0},\nu_{1})$,
so that in this limit, $Q_{s}$ is invariant under $s\mapsto 1-s$ if $S_{0}=S_{1}$.
Recall that $Q_{s}$ has a unique global minimum due to convexity with respect to $s$ \cite{PhysRevLett.98.160501}. Therefore, in the limit $\nu_{0},\nu_{1}\gg 1/2$, the expression (\ref{eqn:altq}) implies that deviations of the symplectic matrices $S_{0}$, $S_{1}$ from each other are responsible for deviations of the critical point from $s=1/2$. In the present analysis, we take $N_{B}\gg 0$ and $\kappa\ll 1$ which together imply $\nu_{0},\nu_{1}\gg 1/2$ and $S_{0}=\mathbb{I}_{1}=S_{1}+O({\kappa N_{S}\over N_{B}})X$, where $X$ is a matrix with $\Vert X\Vert=1$.

Taking $\kappa \ll N_{S}^{-1}$, $N_{B}\gg 0$, and $2N_{S}\ll N_{B}$, we obtain the Bhattacharyya bound
\begin{equation}
{1\over 2}Q_{1/2}^{N} = {1\over 2}\left( 1-{\kappa^{2}(N_{B}-1-2N_{S})\over 8N_{B}} + O\left( {\kappa^{2}N_{S}\over N_{B}^{2}}\right) \right)^{N}
\end{equation}
which is greater than the vacuum transmitter limit of (\ref{eqn:b1}). Therefore, in this parameter domain, optimal detection with a single-mode squeezed transmitter is less useful than optimal detection without illumination. Fig. \ref{fig:ttt}b) shows this phenomenon in a parameter domain with large thermal noise. By contrast, if one considers $\kappa \ll N_{S}^{-1}$, $N_{B}\gg 0$, and $N_{B}<N_{S}$ the Bhattacharyya bound becomes
\begin{equation}
{1\over 2}Q_{1/2}^{N} = {1\over 2}\left( 1-{\kappa^{2}(N_{B}-1)\over 8N_{B}} -{\kappa^{2}N_{S}(N_{S}-N_{B})\over 4N_{B}^{2}}+O\left( {\kappa^{2}N_{S}\over N_{B}^{2}}\right) \right)^{N}
\end{equation}
which is less than the optimal error probability of detection without illumination.

It turns out that for $N_{B}\ll 1$, the value of $Q_{s}$ for a squeezed state transmitter is not generally minimized near $s=1/2$, so the Bhattacharyya bound does not accurately represent the optimal error. Numerically optimizing $Q_{s}$ over $s$ and comparing to (\ref{eqn:b2}), one finds that the ratio of error probability  between the squeezed state transmitter and coherent state transmitter is monotonically increasing with transmitter intensity $N_{S}$. One concludes that single-mode squeezing is not a resource for advantage in QI for $N_{B}\ll 1$, including optical QI.

\section{Not even 6 dB\label{sec:sss}}

Symmetries of the target detection problem imply that it suffices to consider two-mode CV states of the form $\sum_{n=0}^{\infty}c_{n}\ket{n}_{T}\otimes \ket{n}_{Q}$ for optimal QI \cite{PhysRevA.103.062413,Sharma_2018}. Restricting to Gaussian states puts one on the $U(1)\times U(1)$ orbit of two-mode squeezed states (TMSS).
The TMSS of the transmitter and quantum memory is defined by
\begin{equation}
\ket{\psi}_{TQ}={1\over \sqrt{N_{S}+1}}\sum_{n=0}^{\infty}\left( N_{S}\over N_{S}+1\right)^{n\over 2}\ket{n}_{T}\ket{n}_{Q},
\label{eqn:tmss}
\end{equation} with this parametrization chosen so that $\langle a_{T}^{\dagger}a_{T}\rangle = N_{S}=\langle a_{Q}^{\dagger}a_{Q}\rangle$. The covariance matrix $\Sigma_{\rho_{\kappa}}$ of the two-mode state obtained by reflection of the $T$ mode in a thermal environment is equal to the upper $4\times 4$ block of the full $6\times 6$ covariance matrix $\Sigma_{U_{\kappa}^{TE}\ket{\psi}\bra{\psi}_{TQ}\otimes (\rho_{\beta})_{E}U_{\kappa}^{TE\dagger}}$ with matrix elements given by
\begin{align}
(1,1),(2,2) &{:} \; \kappa(2N_{S}+1)+(1-\kappa)(2N_{B}+1)\nonumber \\
(1,2),(1,4),(1,6),(2,3),(2,5),(3,4),(3,6),(4,5),(5,6)&{:} \; 0\nonumber \\
(1,3)&{:} \;  2\sqrt{N_{S}\kappa(N_{S}+1)} \nonumber \\
(1,5)&{:} \; 2\sqrt{\kappa(1-\kappa)}(N_{S}-N_{B}) \nonumber \\
(2,4)&{:}\; -2\sqrt{N_{S}\kappa(N_{S}+1)} \nonumber \\
(2,6)&{:} \; 2\sqrt{\kappa(1-\kappa)}(N_{S}-N_{B}) \nonumber \\
(3,3),(4,4)& \; 2N_{S}+1 \nonumber \\
(3,5)&{:} \; 2\sqrt{N_{S}(1-\kappa)(N_{S}+1)} \nonumber \\
(4,6)&{:} \; -2\sqrt{N_{S}(1-\kappa)(N_{S}+1)} \nonumber \\
(5,5),(6,6) &{:} \; \kappa(2N_{B}+1)+(1-\kappa)(2N_{S}+1) 
\end{align}
%\begin{multline}
% \left(
%  \begin{matrix}
%    \kappa(2N_{S}+1)+(1-\kappa)(2N_{B}+1) & 0 & 2\sqrt{N_{S}\kappa(N_{S}+1)}\\ 
%    0&\kappa(2N_{S}+1)+(1-\kappa)(2N_{B}+1) & 0 \\ 
%    2\sqrt{N_{S}\kappa(N_{S}+1)}&0&2N_{S}+1 \\ 
%    0&-2\sqrt{N_{S}\kappa(N_{S}+1)}&0\\
%    2\sqrt{\kappa(1-\kappa)}(N_{S}-N_{B})&0&2\sqrt{N_{S}(1-\kappa)(N_{S}+1)}\\
%    0&2\sqrt{\kappa(1-\kappa)}(N_{S}-N_{B})&0
%  \end{matrix}\right.                
%\\
%  \left.
%  \begin{matrix}
%    0 & 2\sqrt{\kappa(1-\kappa)}(N_{S}-N_{B})&0 \\ 
%    -2\sqrt{N_{S}\kappa(N_{S}-1)} & 0 & 2\sqrt{\kappa(1-\kappa)}(N_{S}-N_{B})\\
%    0&2\sqrt{N_{S}(1-\kappa)(N_{S}+1)}&0\\
%    2N_{S}+1&0&-2\sqrt{N_{S}(1-\kappa)(N_{S}+1)}\\ 
%    0&\kappa(2N_{B}+1)+(1-\kappa)(2N_{S}+1)&0\\ 
%    -2\sqrt{N_{S}(1-\kappa)(N_{S}+1)}&0&\kappa(2N_{B}+1)+(1-\kappa)(2N_{S}+1)
%  \end{matrix}\right)
%\end{multline}

With $\rho_{\kappa}=\text{tr}_{E}\left[ U_{\kappa}^{TE}\ket{\psi}\bra{\psi}_{TQ}\otimes (\rho_{\beta})_{E}U_{\kappa}^{TE\dagger} \right]$, the symplectic eigenvalues $\gamma_{1}$ and $\gamma_{2}$ of $\Sigma_{\rho_{\kappa}}$ can be obtained using the same method as Ref.\cite{PhysRevLett.101.253601} and for $\kappa \ll 1$ are given by
\begin{align}
\gamma_{1}&= (1 + 2 N_{B}) - {2 (N_{B} (1 + N_{B})) \kappa \over 1 + N_{S} + N_{B}} + o(\kappa) \nonumber \\
\gamma_{2}&= (1 + 2 N_{S}) - {2 (N_{S} (1 + N_{S})) \kappa \over 1 + N_{S} + N_{B}} + o(\kappa).
\end{align}
The first terms in the respective expressions dominate if $\kappa$ is further taken to satisfy $\kappa\ll N_{B}^{-1}$, $\kappa\ll N_{S}^{-1}$, respectively.
In the equation $S\Sigma_{\rho_{\kappa}}S^{T} = \gamma_{1}\mathbb{I}_{2} \oplus \gamma_{2}\mathbb{I}_{2}$, one finds that $S=\mathbb{I}_{4}+O(\sqrt{\kappa})(\mathbb{I}_{2}\otimes Z)$ with $Z=\text{diag}(1,-1)$. The ``target absent'' hypothesis $\rho_{0}$ is now a state on $TQ$ and has covariance matrix $\Sigma_{\rho_{0}}=\beta_{1}\mathbb{I}_{2}\oplus \beta_{2}\mathbb{I}_{2}$ with $\beta_{1}:=2N_{B}+1$, $\beta_{2}:= 2N_{S}+1$ (so $\Sigma_{\rho_{0}}$ is symplectically diagonalized by the identity in $Sp(4,\mathbb{R})$). The structure of the covariance matrix $\Sigma_{\rho_{0}}$ arises from two phenomena: 1. the first two diagonal elements describe total loss of the transmitter mode (the reflected transmitter mode is replaced by the thermal environment), 2. the second two diagonal elements arise from losing all information in the transmitter mode, resulting in a thermal state of the quantum memory $Q$ with $N_{S}$ photons in expectation. In full, the ``target absent'' hypothesis corresponds to the quantum channel $\rho_{\beta}\otimes \text{tr}_{T}$ mapping the set of states of $TQ$ to itself.

\begin{figure}[t]
    \centering
    \includegraphics[scale=0.8]{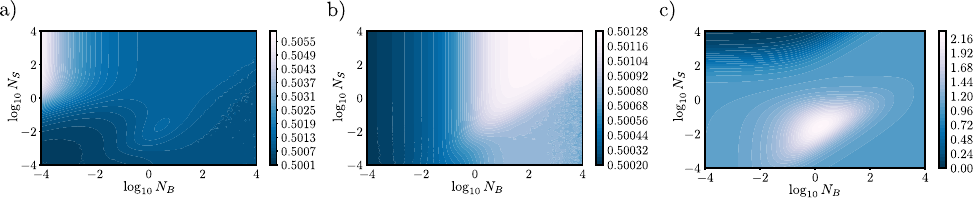}
    \caption{a) Optimal $s$ for $\kappa=10^{-2}$ in (\ref{eqn:uuu}). b) Optimal $s$ for $\kappa=10^{-2}$ for TMSS transmitter. c) ${\xi^{\text{(TMSS)}}\over \xi^{\text{(coherent)}}}$ for $\kappa=10^{-2}$. Note that larger Chernoff exponent $\xi$ indicates smaller error probability in target detection. }
    \label{fig:ppp}
\end{figure} 

Because the 6 dB advantage discussed in Ref.\cite{PhysRevLett.101.253601} is obtained in parameter domain $N_{S}\ll 1$, $N_{B}\gg 1$, $\kappa \ll 1$, we first note that this domain corresponds to $\beta_{1},\gamma_{1} \gg 1$ and $\beta_{2},\gamma_{2}\approx 1$ and symplectic diagonalizations that differ by $O(\sqrt{\kappa})$, so that the rigorous justification of using the Bhattacharyya bound proceeds very similarly to the analysis of (\ref{eqn:altq}). To compare the Bhattacharyya bound of the TMSS transmitter (\ref{eqn:tmss}) to the coherent state result (\ref{eqn:b1}) with equal $N_{S}$ in both cases, we compute $Q_{1/2}$ for the TMSS transmitter to be
\begin{align}
Q_{1/2}^{\text{(TMSS)}}&=1-{(N_{S}-2N_{S}^{3/2}+3N_{S}^{2})\kappa \over N_{B}} - {(N_{B}-1)\kappa^{2}\over 8N_{B}}\nonumber \\
&- {({5\over 4}N_{S}-3N_{S}^{3/2}+9N_{S}^{2})\kappa^{2}\over N_{B}}+O(\kappa^{3})+O(N_{B}^{-3/2})+O(N_{S}^{5/2})
\end{align}
where we kept the $O(\kappa^{2})$ contribution because the vacuum contribution to (\ref{eqn:b1}) is of that order. In terms of the ratio of the Bhattacharyya approximations to the quantum Chernoff exponents, we compare 
\begin{equation}
{\log 2p_{\text{err}}^{\text{(TMSS)}}\over \log 2p_{\text{err}}^{\text{(coherent)}}} \overset{N\rightarrow \infty}{\sim} { {(N_{S}-2N_{S}^{3/2}+3N_{S}^{2})\over N_{B}}+{(N_{B}-1)\kappa\over 8N_{B}} + O\left( {N_{S}\kappa^{2}\over N_{B}}\right) \over {N_{S}\over 2((2-\kappa)N_{B}+1)}+{(N_{B}-1)\kappa\over 8N_{B}}+O\left( {N_{S}\kappa\over N_{B}^{2}}\right)}
\label{eqn:smallnsmallk}
\end{equation}

The $N_{B}\gg 0$ asymptotic shown in (\ref{eqn:smallnsmallk}) is not a continuous function of $(\kappa,N_{S})$ at $(0,0)$, and the 6 dB advantage of the TMSS transmitter (i.e., the factor of 4 in the ratio of the quantum Chernoff exponents) is obtained on paths associated with the limit $\lim_{N_{S}\rightarrow 0}\lim_{\kappa\rightarrow 0}$. This order of limits gives an unachievable value because the target detection problem is not defined for $\kappa=0$. By contrast, there is no advantage on the $\lim_{\kappa\rightarrow 0}\lim_{N_{S}\rightarrow 0}$ paths. For $N_{S}$ going to a fixed positive intensity, as for a realistic transmitter, followed by $\lim_{\kappa\rightarrow 0}$, the advantage is a factor strictly less than 4. Under the model of the target which corresponds to making the substitution $N_{B}\mapsto {N_{B}\over 1-\kappa}$, the ratio of the Bhattacharyya approximations to the quantum Chernoff exponents for the TMSS transmitter and coherent state transmitter is a continuous function of $N_{S}$ and $\kappa$ when $N_{B}\gg 0$, the limit being exactly 4.

For completeness, we show in Fig.\ref{fig:ppp} the numerically identified critical $s$ value over orders of magnitude in both $N_{S}$ and $N_{B}$ for a fixed $\kappa=10^{-2}$. Numerical minimization of $Q_{s}$ becomes challenging for small $\kappa$ because the function becomes constant with respect to $s$. For the same $\kappa$, Fig.\ref{fig:ppp} also shows a maximal factor of 2.23 advantage (3.48 dB) of the TMSS transmitter over the coherent state transmitter, quantified by the ratio of the Chernoff exponents $\xi$ for the respective transmitters. For $N_{B}\ll 1$, closer to the parameter regime of QI at daytime optical frequencies, one finds a domain of transmitter intensities ($N_{S}\gg 1$) for which the TMSS transmitter is disadvantageous compared to coherent state. This conclusion can also be arrived at by carrying out an $N_{B}\ll 1$, $\kappa \ll 1$ expansion of $Q_{1/2}^{\text{(TMSS)}}$ and comparing to (\ref{eqn:b2}).  Similar to the result in Fig.\ref{fig:ttt} in which the increased error probability of a single-mode squeezed vacuum transmitter relative to vacuum state transmitter in a parameter domain was corroborated by a corresponding increased fidelity of the alternatives $\rho_{\kappa}$ and $\rho_{0}$ in that parameter domain, one finds that the parameter domain for a disadvantageous TMSS transmitter is also concomitant with the increased fidelity of $\rho_{\kappa}$ and $\rho_{0}$ compared to the fidelity of the alternatives in the coherent state transmitter QI problem.

\section{Discussion}

Carrying out the thermal environment rescaling $N_{B}\mapsto {N_{B}\over 1-\kappa}$ can change the optimal transmitter and detection strategies for Gaussian QI, not only for transmitters entangled with a quantum memory, but also for single-mode transmitters. When analyzing $p_{\text{err}}$ for single-mode Gaussian QI without this rescaling, we have carefully restricted to parameter regimes in which the Bhattacharyya bound on $p_{\text{err}}$ is a good approximation, then compared squeezed transmitters to vacuum transmitters, finding that for large thermal environment noise and small transmitter intensity, a squeezed transmitter is detrimental for detection compared to passively detecting the target with ``no illumination''. The result is supported by numerical calculations of the Chernoff exponent.

By comparing the Bhattacharyya bounds for a TMSS transmitter and a coherent state transmitter of the same energy, again dispensing with the $N_{B}\mapsto {N_{B}\over 1-\kappa}$, we obtain a value for the gain that is discontinuous at $(\kappa,N_{S})=(0,0)$. Taking the $\kappa \rightarrow 0$ limit for various fixed values of $N_{S}$ determines the set of achievable limiting values of the gain. In particular, 6 dB is a strict upper bound on this set of values. Like Nair's ``no-go'' theorem, which determines the maximal gain factor of non-classical transmitters over coherent state transmitters for noiseless QI ($N_{B}=0$), the present result provides achievable maximal gain factors of the manifold of Gaussian state transmitters for QI in an environment with high thermal background ($N_{B}\gg 0$).

\section*{Acknowledgements}
The author thanks N. Dallmann, R. Newell, K. Meier, D. Dalvit, and P. Milonni for discussions, and acknowledges the LDRD program at Los Alamos National Laboratory. Los Alamos National Laboratory is managed by Triad National Security, LLC, for the National Nuclear Security Administration of the U.S. Department of Energy under Contract No. 89233218CNA000001.

%\printbibliography
\bibliographystyle{unsrt}
\bibliography{np}

\end{document}